\documentclass{elsart}
\usepackage{graphicx}
\usepackage{amssymb}

\begin{document}

\begin{frontmatter}

\title{Evaluation of a long-time temperature drift
in a commercial Quantum Design MPMS SQUID magnetometer using Gd$_2$O$_3$ as a standard}

\author{Sergey L. Bud'ko and Paul C. Canfield}
\address{Ames Laboratory US DOE and Department of Physics and Astronomy, Iowa State University,
Ames, Iowa 50011, USA}

\begin{abstract}
The long-time temperature drift in a commercial Quantum Design MPMS SQUID magnetometer was evaluated using
time-dependent magnetization measurements of Gd$_2$O$_3$. In contrast to earlier claims, the amplitude of the
drift was found not to exceed 1-1.5 K. 30 minutes after system stabilization the temperature deviation did not
exceed 0.2 K and the temperature was fully stabilized in less than 3 hours.
\end{abstract}

\begin{keyword}

magnetic relaxation \sep MPMS magnetometer \sep temperature drift

\PACS 07.55.Jg \sep 74.25.Ha \sep 75.50.Lk
\end{keyword}
\end{frontmatter}

\section{Introduction}
Within last two decades commercial superconducting quantum interference device (SQUID) magnetometers (in a
majority of cases Quantum Design MPMS instruments) have truly become the workhorses of many condensed matter
physics, chemistry and materials science research and educational laboratories. These instruments are widely used
for routine, albeit very accurate, magnetic measurements of wide classes of materials. In the cases of type-II
superconductors in the intermediate state or spin-glasses one of the measurements of choice, allowing deeper
insight into physical processes, is a time-dependent (relaxation) magnetization measurement
\cite{aku02a,cho03a,rav04a,fis99a}. Since a single DC magnetization measurement takes $\sim 30-40$ sec, a long
total measurement time (several hours) is usually required to have a reliable estimate of the relaxation
parameters. In such measurements a knowledge of the long-time temperature drift of the system is important. An
estimate of the long-time temperature drift in Quantum Design MPMS-5 instrument was reported by Kopelevich and
Moehlecke \cite{kop95a}. They reported temperature changes as high as $\sim 4$ K (in 100 - 120 K temperature
range) during $\sim 8$ hours after the temperature was "stabilized" as indicated by the magnetometer's own
thermometer. Whereas \cite{kop95a} was apparently the first publication to address the temperature drift issue in
MPMS instrument, and it sent an important message for the part of the community doing relaxation measurements
using this type of magnetometer, the experimental details in \cite{kop95a} were significantly different from the
case of routine magnetization measurements, potentially affecting the sample thermalization: the temperature was
measured by Pt thermometer mounted in place of a sample. Such a thermometer would, at minimum, require bringing
wires to the sample chamber if magnetization probe was utilized or using Quantum Design {\it Manual Insertion
Utility Probe} (that has a thermal mass significantly higher than the magnetization probe) or its home-made
analogue.

In this work we address the issue of long-time temperature drift of a MPMS instrument using time-dependent DC
magnetization measurements on Gd$_2$O$_3$, a very stable oxide with well-documented \cite{wil18a,wil19a}
Curie-Weiss-like temperature-dependent susceptibility that is known to be used for calibration of susceptibility
measurements \cite{sag03a}. The use of magnetization measurements for temperature-drift evaluation allows for the
experimental protocol to be very similar to routine relaxation measurements and does not alter the thermal mass of
the sample assembly.

\section{Experimental}
\begin{figure}
\begin{center}
\includegraphics[angle=0,width=120mm]{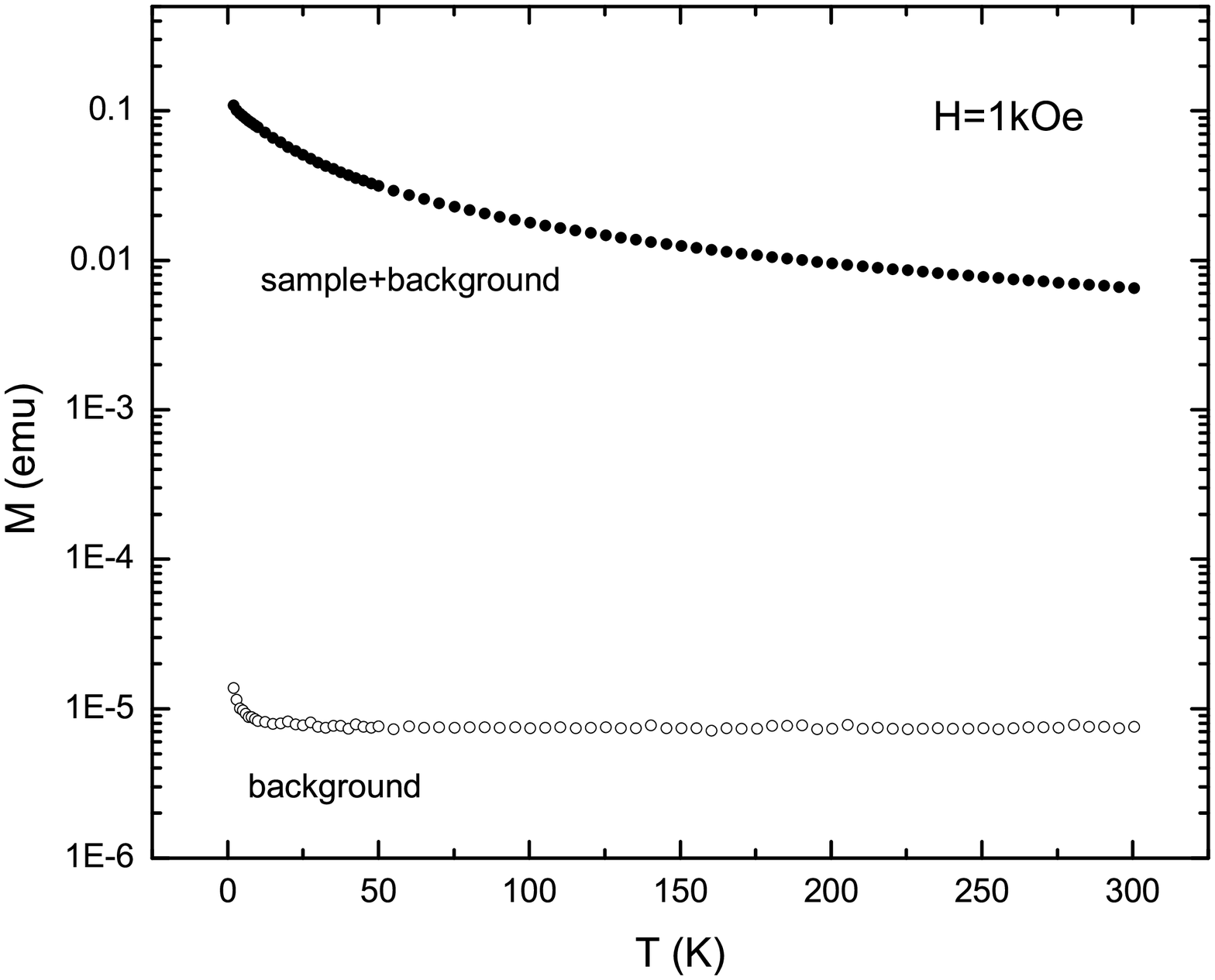}
\end{center}
\caption{\label{f1}Temperature-dependent magnetization of capsule + cotton (background) and the Gd$_2$O$_3$ sample
placed in the same capsule and secured by cotton (sample + background) measured in $H = 1$ kOe applied field.}
\end{figure}

Gd$_2$O$_3$ powder was obtained from the Materials Preparation Center at Ames Laboratory. The highest magnetic
impurities were estimated to be Eu ($\sim 40$ ppm) and Fe ($< 10$ ppm) with total estimated concentration of
impurity elements $< 120$ ppm. To remove possible moisture from the long-time storage, the powder was heated to
$400^\circ$ C for 4 days, cooled down and immediately packed for measurements. The magnetization measurements were
made in a Quantum Design MPMS-7 DC magnetometer equipped with a {\it Continuous Low Temperature Control and
Enhanced Thermometry} (CLTC and ETC) option (M-140) \cite{qdw05a} (the scan length was 6 cm, single measurement
per point was taken, applied field was $H = 1$ kOe). 48.3 mg of loose Gd$_2$O$_3$ powder (the amount was chosen to
ensure reliable signal within the magnetometer's range at 2 K $\leq T \leq$ 300 K) was placed in a gelatine
capsule \cite{gel00a}. The remaining capsule volume was filled with $\sim 50$ mg of cotton so as to reduce
possible motion of the Gd$_2$O$_3$ powder. The capsule with the sample was mounted in a transparent drinking straw
\cite{qda00a}. Temperature dependent magnetization of the capsule with cotton (background) was measured
separately. Over the whole temperature range of interest it was about three orders of magnitude, or more, smaller
than total magnetization of the sample and the capsule/cotton together (Fig. \ref{f1}). Such accuracy is
sufficient for the purpose of this study and in the following, unless stated otherwise, we will refer to the
measured magnetization as the magnetization of the sample.

\begin{figure}
\begin{center}
\includegraphics[angle=0,width=120mm]{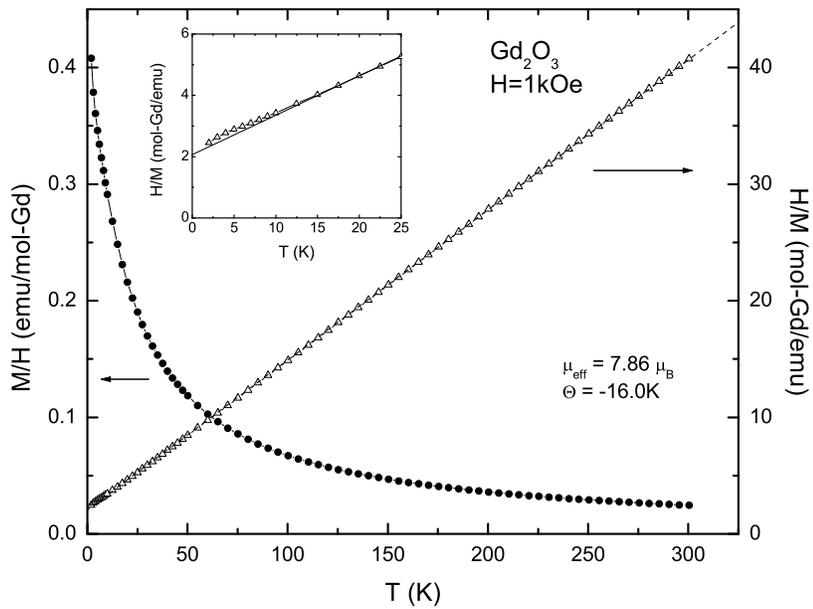}
\end{center}
\caption{\label{f2}Temperature-dependent susceptibility, $M/H$, and inverse susceptibility, $H/M$ of Gd$_2$O$_3$
plotted per mole of Gd$^{3+}$. Dashed line - linear fit to the inverse susceptibility. Inset: low temperature part
of the inverse susceptibility (symbols) with an extrapolation of the linear fit to the high temperature
susceptibility shown as a line.}
\end{figure}

Our temperature-dependent measurements of Gd$_2$O$_3$ show a very clear Curie-Weiss-like behavior (with very small
deviations below $\sim 10$ K) (Fig. \ref{f2}). These data were taken after stabilizing at 50 K, cooling down to 2
K and taking data on warming, stabilizing temperature every 1 K up to 10 K, every 2.5 K up to 50 K and every 5 K
up to 300 K. From the linear fit of the inverse susceptibility, $\chi^{-1} = H/M$, the Gd$^{3+}$ effective moment
was estimated to be $\mu_{eff} = 7.86 \pm 0.05 \mu_B$, close to the theoretical value of 7.94 $\mu_B$ for the
Gd$^{3+}$, with the Curie-Weiss temperature of $\Theta = -16.0$ K, similar to the literature data
\cite{wil18a,wil19a,sag03a,moo75a}.

To analyze the long-time temperature drift of the instrument we will use the Curie-Weiss behavior of the
susceptibility of Gd$_2$O$_3$. Since (at least between 10 K and 300 K) the existing data and the current
understanding of the magnetic properties of Gd$_2$O$_3$ give no indication of magnetic relaxation processes
related to the material, we will assume that any change in the measured magnetization at a nominally fixed
temperature is caused by a long-time temperature drift of the sample. To avoid processes related to relaxation of
the magnetic field in the superconducting magnet of the instrument, the field was set to 1 kOe in the persistent
mode about an hour before the time dependent measurements were started and was not changed during the
measurements. We will not use our $\chi(T)$ data to extract the exact temperature of the sample during such
measurements, but rather will look at the evolution of the magnetization and will convert the magnetization drift
$\Delta M$ (taken with respect to the "stabilized" magnetization after $\sim 10$ h of measurements) into a
temperature drift $\Delta T$. It should be noted that for convenience in dealing with the experimental data,
simple equations below are written for the sample as it is, without a normalization to its mass or molecular
weight. The interested reader can use the sample mass and the molar mass to convert these equations into whatever
form or units are convenient. We can write the inverse magnetization as

\begin{equation}
\label{eq1} 1/M = a + bT
\end{equation}

The parameters $a$ and $b$ are determined from the linear fit of the inverse magnetization. From the Eq. \ref{eq1}
we obtain

\begin{equation}
\label{eq2} \frac{dM}{dT} = - \frac{b}{(a+bT)^2} = - \frac{b}{M^2}
\end{equation}

Consequently

\begin{equation}
\label{eq3} \Delta T = \Delta M \frac{1}{dM/dT} =  - \Delta M  \frac{M^2}{b}
\end{equation}

From the $M(T)$ measurements in $H = 1$ kOe the value of $b$ was determined to be 0.4828 emu$^{-1}$K$^{-1}$.

Two sets of time-dependent magnetization measurements were performed. For each measurement temperature in the
first set the sample was heated to 300 K, held there for 10 sec and then cooled to the desired temperature at
nominal 10 K/min rate. 10 sec after the temperature was "stabilized" as determined by the magnetometer's
thermometer the time-dependent measurements started. For the second set the sample was first cooled down to 10 K,
kept at that temperature for 10 sec and then heated up to the desired temperature at nominal 10 K/min rate.
Similarly to the first set, 10 sec waiting time was included before the start of the time-dependent measurements.
An example (for $T_{set} = 50$ K in the first set of data) of the sequence used for time-dependent magnetization
measurements is given below:
\\

\begin{ttfamily}
{\scriptsize Set Temperature 300.000K at 10.000K/min}
\newline {\scriptsize Waitfor Temp:Stable Delay:10secs}
\newline {\scriptsize Set Temperature 50.000K at 10.000K/min}
\newline {\scriptsize Waitfor Temp:Stable Delay:10secs}
\newline {\scriptsize Scan Temp from 50.00K to 50.00K at 3.000K/min in 200 steps (0K/step) Settle}
\newline {\scriptsize Waitfor Temp:Stable Delay:30secs}
\newline {\scriptsize Measure DC: 6.00cm, 48pts, 1scans, AutoRng, Long, Iterative Reg., track:Yes, raw:No, diag:No}
\newline {\scriptsize End Scan}
\end{ttfamily}

\section{Results}
\begin{figure}
\begin{center}
\includegraphics[angle=0,width=100mm]{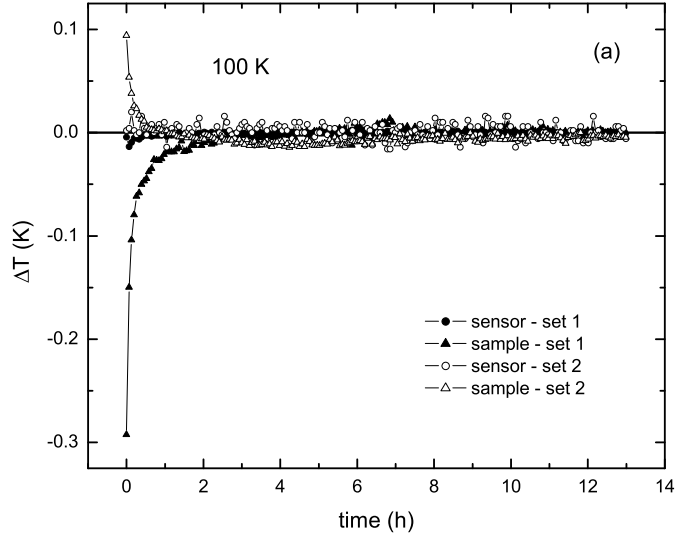}
\includegraphics[angle=0,width=100mm]{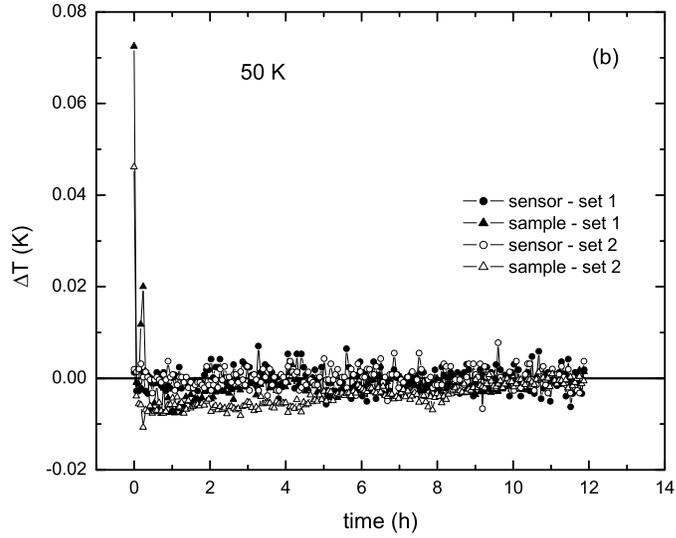}
\end{center}
\caption{\label{f3}Example of time-dependent temperature drift for (a) $T_{set} = 100$ K and (b) $T_{set} = 50$ K
as measured by the magnetometer's sensor and as inferred from the changes in measured magnetization of Gd$_2$O$_3$
for two sets of measurements (see text for details).}
\end{figure}

Fig. \ref{f3}(a) shows the time-dependent temperature drift for $T_{set} = 100$ K as measured by the
magnetometer's sensor and as calculated using the Eq. \ref{eq3} from the changes in measured magnetization of
Gd$_2$O$_3$ for two sets of measurements - after heating to 300 K (set 1) and after cooling to 10 K (set 2). In
both cases the magnetometer's thermometer does not show any time drift and these data simply indicate the
temperature noise tolerated by the magnetometer's stability algorithm. The temperature drift as inferred from the
magnetization drift is $\sim 0.3$ K (with temperature increasing with time) in the first set and $\sim 0.1$ K
(with temperature decreasing with time) in the second set. Different signs of $\Delta T$ are consistent with the
protocol of the measurements in each set. In both sets the temperature is practically stable after $\sim 3$ hours
and within 0.1 K (0.1\% accuracy) within less than an hour. For lower temperature, $T_{set} = 50$ K (Fig.
\ref{f3}(b)) the temperature drift is significantly smaller and the temperature is stabilized within $\sim 0.01$ K
in less than an hour.

\begin{figure}
\begin{center}
\includegraphics[angle=0,width=120mm]{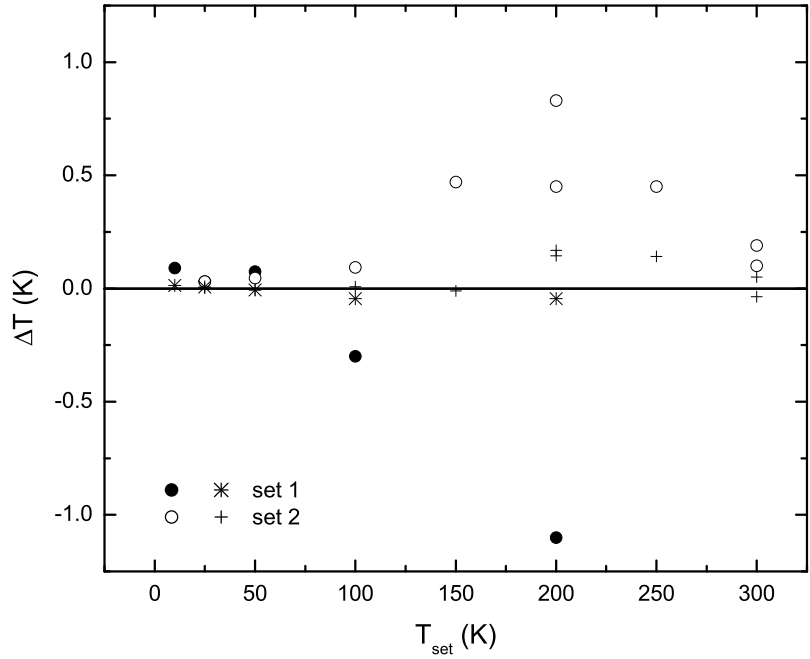}
\end{center}
\caption{\label{f4}($\bullet,\circ$) - Difference between initial temperature (10 sec after the temperature was
"stabilized" according to the magnetometer's sensor) and stabilized temperature as inferred from the
time-dependent magnetization measurements as a function of set-temperature for two different protocols - set 1
(cooling from 300 K) and set 2 (warming from 10 K); ($\ast,+$) - difference between the temperature after 30 min
of measurements and the stabilized temperature.}
\end{figure}

Data for different set temperatures are summarized in Fig. \ref{f4}. Below 100 K the maximum temperature deviation
is less than 0.1 K. At and above 100 K the deviation depends of the cooling/warming history and reaches maximum of
$|\delta T| \sim 1$ K at $T_{set} = 200$ K. These data represent the extreme temperature excursions in our data.
If we plot two similar data sets for the $t = 30$ min data (Fig. \ref{f4}) the deviations from the stabilized
temperatures will be $< 0.2$ K. Over the whole temperature range the temperature is fully stabilized after $t \leq
1$ h below 100 K and after $t \leq 3$ h at and above 100 K. Different time-dependent measurements with nominally
the same history may have some spread of $\Delta T$ ({\it e.g.} 200 K and 300 K for set 2) so the results in Fig.
\ref{f4} serve only as guidelines.

\section{Conclusions}
We evaluated the long-time temperature drift in MPMS SQUID magnetometer between 10 K and 300 K using Gd$_2$O$_3$
time-dependent magnetization measurements and well-established Curie-Weiss-like $M(T)$ behavior of this material
utilizing the measurement schedule similar to that used for magnetic relaxation measurements in superconductors or
magnetic materials. The maximum temperature deviation and its relaxation depend on the measurement's history and
was found to be within 0.1 K below 100 K and higher, up to 1 K above 100 K. Our estimates give lower $|\Delta T|$
values and shorted temperature stabilization times than those reported in Ref. \cite{kop95a}. This may be due to
different experimental approaches for such an estimate. Additionally, our measurements were performed on an
upgraded version of the MPMS instrument, so if the magnetometer in Ref. \cite{kop95a} did not have the CLTC and
ETC option, an improved temperature control in our instrument is probably (at least partially) the reason for its
lower maximum temperature deviations. Our results can serve as a guideline for designing measurement's schedule
for magnetic relaxation studies in commercial Quantum Design MPMS magnetometers.

\ack We thank R.W.McCallum for sharing his stock of Gd$_2$O$_3$ with us. Ames Laboratory is operated for the U.S.
Department of Energy by Iowa State University under Contract No. W-7405-Eng-82. This work was supported by the
Director for Energy Research, Office of Basic Energy Sciences.

\clearpage

\clearpage

\clearpage

\clearpage

\end{document}